\documentclass[aps,prl,twocolumn,groupedaddress,nofootinbib,nobibnotes,showkeys]{revtex4-2}
\usepackage{graphicx}
\usepackage{amssymb}
\usepackage{appendix}
\usepackage{hyperref}
\usepackage{enumitem}
\usepackage{amsmath}

\begin{document}

\title{On the statistical arrow of time}

\author{Andreas Henriksson}
\email[]{andreas.henriksson@skole.rogfk.no}
\affiliation{Stavanger Katedralskole, Haakon VII's gate 4, 4005 Stavanger, Norway}

%\date{\today}

\begin{abstract}

What is the physical origin of the arrow of time? It is a commonly held belief in the physics community that it relates to the increase of entropy as it appears in the statistical interpretation of the second law of thermodynamics. At the same time, the subjective information-theoretical interpretation of probability, and hence entropy, is a standard viewpoint in the foundations of statistical mechanics. In this article, it is argued that the subjective interpretation is incompatible with the philosophical point of view that the arrow of time is a fundamental property of Nature. The subjectivist can only uphold this philosophy if the role played by the second law of thermodynamics in defining time's arrow is abandoned.

\end{abstract}


\maketitle

\subsection{Introduction}

In this article, we will study the arrow of time as it appears in the theory of statistical mechanics. For this purpose, it is necessary to clearly state the premises on which the study is based, regarding the interpretation of probability being adopted. Broadly speaking, the set of probability interpretations can be divided into two groups. The objective interpretation \cite{cramer, feller} regards the probability of an event as an objective property of that event and independent of the observer of the event. On the contrary, in the subjective interpretation \cite{keynes, jeffreys}, probabilities arise due to the observer's lack of complete knowledge, or information, about the event. In other words, the appearance of probabilities arises due to the ignorance of the observer. With the development of information theory \cite{shannon}, the subjectivist philosophy was applied to the foundations of statistical mechanics \cite{jaynes1, jaynes2}. The problem of the arrow of time in statistical mechanics will in this article be addressed from the subjective information-theoretic point of view. It should be remarked that this interpretation is not universally accepted. Furthermore, it should be noted that there are no known physical observations that allow for a distinction between objective and subjective interpretations. The mathematics of statistical mechanics and its predictions are the same for both interpretations. It is then a matter of personal taste which philosophy is adopted. At this stage, it should be emphasized that the subjective point of view is not necessarily reflective of the author's personal view. That view has, frustratingly, changed many times back and forth over the years. The intention of the article is rather to deduce the consequences, concerning the arrow of time, if the subjective viewpoint is taken. We will conclude that there are two possibilities:

\begin{itemize}
\item The arrow of time is not a fundamental property of Nature. It is subjective due to the ignorance of the observer. The apparent directionality to time thus exists only in the minds of ignorant observers.
\item The arrow of time is a fundamental property of Nature. However, its origin has nothing to do with the concept of entropy and the second law of thermodynamics.
\end{itemize}

The organization of the article is as follows. The discussion, from here on, will take place with the subjective glasses put on. Whatever is stated in the article is from the subjective information-theoretic perspective. It is first in the section on the arrow of time that this viewpoint is put into question, by illustrating the seemingly bizarre consequence that a monkey should have a greater sense of time as compared to a human physicist.

To illustrate the concept of entropy, from the philosophical point of view that it originates from the ignorance of the observer, consider the following situation. Consider a gas of molecules that are constrained within a box of small volume in the corner of a room. As a door in the box is opened, the gas molecules will spread out into the room. That the gas spreads outward into the room, eventually filling all of it, rather than staying put in the corner, or coming back to the corner within a measurable passage of time, is argued to define a direction for the arrow of time. When the gas has spread out evenly in the room, the room is in thermal equilibrium with the gas molecules. Thus, it seems to be the case that systems, which are left to themselves, tend to evolve in time in such a way that they reach thermal equilibrium. From the statistical mechanical point of view, thermal equilibrium happens when all possible microstates are equally probable. The directionality in time is in this picture thus viewed as being due to the flow of probability from a non-uniform to a uniform distribution. The concept of entropy then enters as a type of measure of how far this process has come. For uniform probability distributions, the entropy is at its maximum value. The physical interpretation of this situation is that all states being equally probable means that the uncertainty experienced by an observer about the exact degrees of freedom of all gas molecules is at its maximum. Any given pair of molecules can be interchanged without the observer noticing it.

As another pedagogical illustration, consider an ordered stack of cards laid out on a table in an array. The system of cards has a well-defined and clear state that is uniquely defined by the locations of cards on the table. Any disturbance to this state, such as interchanging the positions of a pair of cards, will yield a new state which is distinct from the ordered initial state. Any human observer can easily tell that the cards have changed places. The entropy of the system of cards is at its minimum. Consider now that the initial condition is changed by throwing the set of cards up in the air and letting them fall randomly on the table. There is no clear pattern to how the cards are positioned on the table. Yet, each configuration of cards on the table represents a unique state. If a pair of cards switch places, the physical state does change. However, since the initial configuration of cards on the table was random, with no apparent order, it is much more difficult for an observer to notice the change in the state coming from interchanging the locations of a pair of cards. If all cards are completely randomly scattered in an even fashion on the table, it is said that the entropy for the system of cards is, from the viewpoint of the observer, at its maximum. This means that it is in all practicality impossible for the observer to determine any changes in the state of the system due to the interchanging locations of cards. Whatever the locations of the cards are, and however they are interchanged, the system will look the same from the perspective of the observer. When this is the situation, the observer assigns a uniform probability distribution for the system of cards. All possible configurations of cards are equally probable. Entropy is thus a measure of the uncertainty, or ignorance, possessed by the observer about the physical state of the system.

\subsection{Uncertainty and coarse-graining}

Describing the dynamics of a system is quite complicated. Most systems of interest contain a vast number of particles that interact in complicated ways. For such large systems, it is usually extremely hard to track the individual evolution of each particle as the system evolves in time. The observer does not possess enough information to state with certainty the positions and velocities of all particles. Perfect knowledge about the position and velocity, or momenta, of each particle is lost. The observer loses information about the system over time. It is lost not because of a fundamental violation of information conservation in Nature but merely because of the difficulty for an observer to keep track of all the degrees of freedom. Therefore, from the perspective of the observer, there is an uncertainty $\Delta q$ associated with the position of a state and an uncertainty $\Delta p$ associated with the momentum of a state. For this reason, the observer is unable to determine with absolute certainty the state of the system at any given time. The observer can only determine whether the system occupies a state which lies within any given region $\Omega_j$ on phase space, whose volume $V_{\Omega_j}$ is given by the uncertainties $\Delta q$ and $\Delta p$, i.e.
\begin{equation}
V_{\Omega_j}=\Delta q \Delta p.
\end{equation}
The volume $V_{\Omega_j}$ is thus a measure of how ignorant the observer is about the details of the system, in the sense that the observer cannot locate an individual state to a greater precision than the size of $\Omega_j$. Due to this lack of precision, the observer is unable to distinguish between states that lie within $\Omega_j$. All states within $\Omega_j$, with their sets of degrees of freedom, have, from the perspective of the observer, collapsed into a single state whose single set of degrees of freedom is given by $q+\Delta q$ and $p+\Delta p$. This so-called coarse-grained, or mixed, state is not a fundamental or pure, state of the system. It is a description that averages over all pure states within $\Omega_j$. Put differently, a mixed state $\psi_j$, $j\in [1, M]$, where $M$ is the number of mixed states on phase space, is a subjective representation, by an ignorant observer, of a collection of pure states $\phi_\alpha, \alpha\in[1, N]$, where $N$ is the number of pure states within $\Omega_j$. As the system evolves in time, the observer is only able to measure the coarse-grained flow, i.e., the jumping from one mixed state $\psi_j$ to a different mixed state $\psi_i, i\neq j$.

It should be noted that due to the lack of perfect knowledge about all the relevant degrees of freedom, the observer is unable to predict a unique evolutionary path on phase space along which the system evolves.

\subsection{Probability conservation}

Due to the ignorance of the observer, i.e., the observer's inability to distinguish the set of pure states within any given coarse-grained region $\Omega_j$, it is necessary to introduce the notion of probability on phase space. Let $P_j$ be the probability that the system occupies the region $\Omega_j$ and let $P_\alpha$ be the probability that the system occupies the pure state $\phi_\alpha$ within $\Omega_j$. If the observer knows with absolute certainty that the system occupies the mixed state $\psi_j$ and not some other state $\psi_i, i\neq j\in[1, M]$, it is given that
\begin{eqnarray}
P_i&=&0, \forall i\neq j\in[1, M],\\
P_j &\equiv & \sum_{\alpha=1}^N\ P_\alpha=1.
\end{eqnarray}
For continuous systems, the summation is replaced by an integral, i.e.
\begin{equation}
P_j\equiv\int_{\Omega_j}P_{\alpha}\ dV_{\alpha}=1.
\end{equation}
where $dV_{\alpha}=dq_{\alpha}dp_{\alpha}$ is the phase space volume of the pure state $\phi_{\alpha}$. If the knowledge possessed by the observer about the coarse-grained flow of the system is not lost over time, i.e., information is conserved, then the probability $P_j$ is constant in time, i.e.
\begin{equation}
\frac{dP_j}{dt}=0.
\end{equation}
In other words, it is assumed that there is no loss of probability from $\Omega_j$ to any other coarse-grained region $\Omega_i, \ i\neq j$. 

Written in terms of the probabilities $P_\alpha$, the condition of no loss of coarse-grained knowledge become
\begin{eqnarray}
\frac{dP_j}{dt}&=&\frac{d}{dt}\int_{\Omega_j}P_{\alpha}\ dV_{\alpha}\nonumber\\
&=&\int_{\Omega_j}\left(\frac{dP_\alpha}{dt}+P_\alpha\ \vec{\nabla}\cdot\vec{v}\right)\ dV_{\alpha}\nonumber\\
&=&0,
\end{eqnarray}
where $\vec{v}=\left(\dot{q}, \dot{p}\right)$ define the phase-space velocity of the Hamiltonian flow. Since this should hold independently on the size of $\Omega_j$, the integrand must identically vanish, i.e.
\begin{equation}
\frac{dP_\alpha}{dt}+P_\alpha\ \vec{\nabla}\cdot\vec{v}=0.
\end{equation}
This is the continuity equation for probability flow within any given coarse-grained region $\Omega_j$. It is referred to as the Liouville equation for the probability distribution within $\Omega_j$.

\subsection{Statistical equilibrium}

The continuity equation can be rewritten, showing that probability is locally conserved within $\Omega_j$. Using the total time derivative of $P_\alpha$, i.e.
\begin{equation}
\frac{dP_\alpha}{dt}=\frac{\partial P_\alpha}{\partial t}+\vec{\nabla}P_\alpha\cdot\vec{v}
\end{equation}
and the product rule
\begin{equation}
\vec{\nabla}\cdot\left(P_\alpha\ \vec{v}\right)=\vec{\nabla}P_\alpha\cdot\vec{v}+P_\alpha\ \vec{\nabla}\cdot\vec{v},
\end{equation}
the continuity equation becomes
\begin{equation}
\frac{\partial P_\alpha}{\partial t}+\vec{\nabla}\cdot\left(P_\alpha\ \vec{v}\right)=0.
\end{equation}
The term $\vec{\nabla}\cdot\left(P_\alpha\ \vec{v}\right)$ represents the difference between the probability outflow and inflow for the pure state $\phi_\alpha$.

Consider a system that has been closed for a sufficiently long period such that the density of pure states within $\Omega_j$, and hence $M$, do not change with time. In this situation, the probability distribution $P_\alpha$ has no explicit dependence on time. The continuity equation is then reduced to
\begin{equation}
\label{eq:statisticalequilibrium}
\vec{\nabla}\cdot\left(P_\alpha\ \vec{v}\right)=0.
\end{equation}
This is the mathematical condition the system needs to satisfy for it to be said to exist in statistical equilibrium. In other words, a system is in statistical equilibrium if there is no net probability flow on phase space.

The incompressibility of the Hamiltonian flow implies that the time the system spends in any single pure state, before evolving to the next single pure state, is the same for all pure states. If this were not the case, the state points on phase space would lump together which would signify a violation of Liouville's theorem \cite{henriksson}. This implies that throughout an extended period, the total time spent by the system in any given pure state is expected to be the same for all pure states. This expectation, which is due to a combination of the Liouville theorem and the law of large numbers, is in this article interpreted to be equivalent to the ergodic theorem of statistical mechanics \cite{boltzmann2}\cite{birkhoff}\cite{neumann}. Let $n_\alpha$ denote the number of times the system occupies the pure state $\phi_\alpha$. The total number of times, $n$, the system occupies the set of $N$ pure states within $\Omega_j$ is then
\begin{equation}
n=\sum_{\alpha=1}^N\ n_\alpha.
\end{equation}
The ergodic theorem then says that over an extended period, such that $n$ is large, it is expected that the system occupies all pure states within $\Omega_j$ an equal number of times, i.e.
\begin{equation}
n_\alpha=n_\beta , \ \forall \beta\neq\alpha\in[1, N],
\end{equation}
such that
\begin{equation}
n=N\cdot n_{\alpha}.
\end{equation}

It is now possible to define the notion of a probability $P_\alpha$ for the pure state $\phi_\alpha$ of a closed system from the notion of a relative frequency,
\begin{equation}
\label{eq:microcanonicalprobability}
P_\alpha\equiv \lim_{n\rightarrow\infty}\frac{n_\alpha}{n}=\frac{n_\alpha}{N\cdot n_\alpha}=\frac{1}{N}.
\end{equation}
Thus, all the pure states within $\Omega_j$ are equally probable. This implies that an observer has lost all information, down to the scale of $V_{\Omega_j}$, about the system, since no distinctions can be made between the possible pure states within $\Omega_j$. The uniform probability distribution given by equation \ref{eq:microcanonicalprobability} is commonly referred to as the microcanonical \cite{gibbs}, or fundamental \cite{khintchine}, probability distribution. Thus, given that the system satisfies the Liouville theorem, the microcanonical probability distribution satisfies the condition for statistical equilibrium.

\subsection{Ergodicity breaking}

There exist also non-uniform probability distributions. The non-uniformity arises due to interactions that the system has or has had in the not-too-far-distant past with an environment. In other words, the system is, or was recently, not isolated. Due to the interaction with an environment, the density of states changes with time. If the interaction is uniform on phase space, the density changes uniformly on phase space. However, in general, this is not the case. An interaction, characterized by potential energy, does depend on the specific values for the generalized coordinates. In that scenario, the density of states is a local function of phase space. This has the consequence that the total time spent by the system within any given region on phase space is not necessarily the same as within any other equally sized region. In other words, the ergodic theorem appears to be violated. Thus, not only is the probability distribution non-uniform when there is a non-negligible net interaction with the environment, but it can also change over time. To put it differently, if there exists an interaction between the system and its environment, as seen from the perspective of an observer of the system, this implies that the observer possesses the knowledge, i.e., information, about the interaction. This information is used by the observer when assigning probabilities for the possible states of the system. The fact that the observer possesses some amount of information means necessarily that the probability distribution is non-uniform. It is only at statistical equilibrium, where all information is lost, that the observer assigns a uniform probability distribution.

From the definition of probability in statistical equilibrium it is clear that the probability for any given pure state decrease as the number of pure states $N$ increase, i.e., as the uncertainty volume increase. In non-equilibrium, where probabilities are not equal, it is the average probability that decreases as the uncertainty volume increase.

It should be emphasized that the apparent violation of the ergodic theorem is not of a fundamental character. It is only because the degrees of freedom associated with the environment cannot be excluded when defining the degrees of freedom for the system. In other words, the environment should be included in the definition of the system. If that is done then there exists no environment and hence there cannot be any net transfer of energy and particles from, or to, the system. Then, this redefined system, which considers all degrees of freedom, even those which the experimenter may think to belong to an 'environment', do indeed conserve information, and ergodicity is not broken. The probability distribution for the states of this redefined system is uniform, i.e., all mixed states for any given system, assuming the system has been defined such that no degrees of freedom are being forgotten, are equally probable. In most practical situations, however, an environment for any system under study will always exist. The question is to what degree this environment interacts with the system. The weaker the interaction, the weaker the ergodicity breaking, and the closer will the system come to a uniform probability distribution.

\subsection{Entropy}

A measure for the amount of information possessed by the observer, i.e., the amount of uncertainty in the determination of the pure state of the system, should depend on the probability distribution $\left\{ P_\alpha\right\}$. This measure is denoted by $S(\left\{ P_\alpha\right\})$ and referred to as the entropy of the system. To obtain a specific form for the entropy as a function of the probability distribution, it is noted that this function should satisfy the following conditions.

\begin{enumerate}[label=\roman*]
\item The entropy should be zero when the observer has complete knowledge about the evolution of the system. In other words, if the observer knows with absolute certainty that the system occupies a specific state $\phi_\alpha$, such that $P_\alpha=1$ and $P_\beta=0\ \forall \beta\neq\alpha$, the entropy must vanish.
\item The entropy should always be either zero or a positive number, i.e. $S\geq 0$.
\item The entropy should take a maximum value when the observer is maximally ignorant. This happens when the system is in statistical equilibrium. When all states are equally probable, it implies that the observer possesses zero partial knowledge which can be used to distinguish between some of the features of the set of states. Thus,
\begin{equation}
P_\alpha=\frac{1}{N}\ \forall\alpha\in[1,N]\ \rightarrow S(\left\{ P_\alpha\right\})=S_{max}.
\end{equation}
\item The entropy should, in statistical equilibrium, be a continuously increasing function of the number of states $N$. In other words, when $N$ increases, the uncertainty volume  $V_{\Omega_j}$ increases continuously.
\item The entropy should satisfy the following composition law,
\begin{equation}
S(\left\{ P_\alpha\right\}\cdot\left\{ P_\beta\right\})=S(\left\{ P_\alpha\right\})+S(\left\{ P_\beta\right\}).
\end{equation}
This composition law is understood as follows. Let $\Omega_j$ be divided into two subregions $\Omega_j^\alpha$ and $\Omega_j^\beta$ such that $V_{\Omega_j}=V_{\Omega_j^\alpha}+V_{\Omega_j^\beta}$. The states $\phi_\alpha, \alpha\in[1, N_\alpha]$, belong to $\Omega_j^\alpha$ and the states $\phi_\beta, \beta\in[1, N_\beta]$, belong to $\Omega_j^\beta$, where $N_\alpha+N_\beta=N$. The corresponding probability distributions, $\left\{ P_\alpha\right\}_{\alpha=1}^{N_\alpha}$ and $\left\{ P_\beta\right\}_{\beta=1}^{N_\beta}$, satisfy $\sum_{\alpha=1}^{N_\alpha}\ P_\alpha+\sum_{\beta=1}^{N_\beta}\ P_\beta=1$ and, due to them being independent of each other, their product give the probability distribution associated with the region $\Omega_j$, i.e. $P(\Omega_j)=\left\{ P_\alpha\right\}\cdot\left\{ P_\beta\right\}$. The composition law thus states that the total uncertainty within region $\Omega_j$ is the sum of the uncertainties associated with the subregions of $\Omega_j$.
\end{enumerate}
Conditions (i) and (v) suggest that the entropy has a logarithmic dependence on the probability distribution. Condition (ii) suggest that it is necessary to include an additional minus sign in the definition of entropy. This is seen from the general definition of $P_\alpha$, i.e.
\begin{equation}
\log{P_\alpha}=\lim_{n\rightarrow\infty}\log\left(\frac{n_\alpha}{n}\right)=\log{n_\alpha}-\lim_{n\rightarrow\infty}\log{n}<0,
\end{equation}
which, for a system in statistical equilibrium become
\begin{equation}
\log{P_\alpha}=\log{\frac{1}{N}}=\log{1}-\log{N}=-\log{N}<0.
\end{equation}
Since the entropy function should act as a measure for systems both in and out of statistical equilibrium, i.e., for both uniform and non-uniform probability distributions, it is required to take the statistical average of all logarithmic contributions to the entropy, i.e.
\begin{eqnarray}
S(\left\{ P_\alpha\right\})&\sim& -\left(\frac{n_1}{n}\log{P_{1}}+\cdot\cdot\cdot +\frac{n_N}{n}\log{P_{N}}\right)\noindent\\
&\sim&-\sum_{\alpha=1}^N\ \frac{n_\alpha}{n}\ \log{P_\alpha}\noindent\\
&\sim&-\sum_{\alpha=1}^N\ P_\alpha\log{P_\alpha}.
\end{eqnarray}
This entropy function then satisfies conditions (iii) and (iv). With the proportionality constant identified with the Boltzmann constant $k_B$, it is referred to as the Gibbs entropy \cite{gibbs} and is, in the information-theoretic language, identical to the Shannon entropy \cite{shannon}\cite{jaynes1}\cite{jaynes2}.

In conclusion, the entropy of a system measures the amount of information within the system, and it is given by the Gibbs formula
\begin{equation}
S(\left\{ P_\alpha\right\})=-k_B\sum_{\alpha=1}^N\ P_\alpha\log{P_\alpha}.
\end{equation}
In statistical equilibrium, the Gibbs entropy reduces to the Boltzmann entropy \cite{boltzmann2}\cite{boltzmann1},
\begin{equation}
S=k_B\log{M}.
\end{equation}
It is important to emphasize that entropy, within the subjective interpretation, is not a physical quantity in the same manner as e.g. energy. It is determined by the probability distribution of the states of the system assigned by the observer and as such, it is a quantity that depends both on the specifics of the system and the amount of information possessed by the observer.

\subsection{Second law of thermodynamics}

\begin{figure}[h!]
 \centering
  \includegraphics[width=0.45\textwidth]{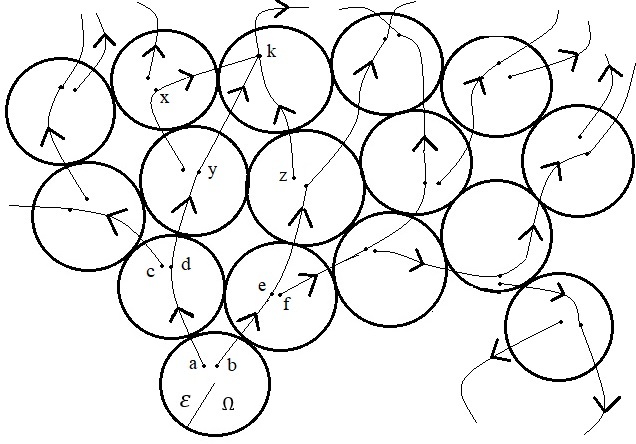}
 \caption{Irreversible, entropy increasing, flow on phase space.}
 \label{fig:Fig1}
 \end{figure}

If the state of a system is known with infinite precision at some given time, and if the laws of motion are known to infinite precision, then any earlier or later states of the system can be predicted with infinite precision. In such a deterministic situation, information about the system is never lost. However, in practical reality, the experimental precision by which the state can be determined is limited. Instead of knowing the initial conditions with infinite precision, they are known to some degree of error, $\epsilon$, on phase space. Therefore, the state of the system is only known to lie within a finite region, $\Omega$, of radius $\epsilon$ and volume $V_{\Omega}$. As the system evolves from the initial conditions, it is impossible to predict the exact path of phase space. Any two neighboring states within $\Omega$, e.g. $a$ and $b$, see Fig. \ref{fig:Fig1}, might evolve differently over time. 
State $a$ might evolve into either state $c$ or state $d$. Due to the limited precision, it is impossible to say which state it evolve into. State $b$, on the other hand, might evolve into state $e$ or state $f$. This process of diverging paths continues as time unfolds. Therefore, the number of states in which the system might exist increases over time. In other words, the amount of uncertainty, i.e., the entropy, increases with time. Alternatively put, over time, any observer will continue to lose information about the system because of not knowing the initial conditions of the system with infinite precision. It is also possible for the entropy to decrease over time meaning that the observer has gained information about the system. This corresponds to the situation when possible paths converge at some point. For example, the states $x$, $y$ and $z$ all converge into state $k$. The uncertainty of the system has thus decreased since there are now fewer possible states in which the system might exist. However, the probability that paths converge to a single state is much lower than the probability that they diverge to separate states. The reason for this is that the state $k$ is merely one possible state out of a large number of possible states within volume $V_{\Omega}$ which $x$, $y$ and $z$ could have evolved into. Thus, overall, the observer loses information exponentially over time. Eventually, all information has been lost. The observer has become maximally ignorant. The entropy has reached its maximum value. At this stage, the system has reached statistical equilibrium where all states are equally probable since the observer is unable to make any distinctions between them. This tendency, of any given system, as viewed by an observer with limited knowledge of the initial conditions, to increase its entropy and evolve toward statistical equilibrium, is referred to as the second law of thermodynamics.

It is important to emphasize that the apparent violation of determinism and reversibility, i.e., violation of the Liouville theorem, is from the viewpoint of the subjective information-theoretic interpretation of probability not due to a fundamental character in the dynamical evolution of systems. The apparent irreversibility is due to the ignorance of the observer.

\subsection{Arrow of time}

If the point of view is that it is the second law that dictates the directionality of time, then the following conclusion must follow: For an infinitely wise observer, who can determine the initial conditions and the laws of motion with infinite precision, the evolution of the system is completely reversible in time. For such an enlightened observer, there is no arrow of time. Time does not flow into the future from the past. The apparent unique direction in which time flow, i.e., toward the future, is merely a consequence of the fact that the observer does not possess infinite knowledge about the system under consideration. For such an ignorant observer, it is exponentially more probable that the system evolves in such a way that possible paths diverge on phase space. The diverging evolution defines the direction, or arrow, of time as seen from the perspective of the observer. In the unlikely scenario that the possible paths converged at a quicker rate than they diverged, such that information on average was gained, then the system would be observed to evolve backward in time. The logical philosophical question to ask is then the following:\\
\\
\textit{Can the seemingly universal feature of the arrow of time, flowing towards the future for all observers, really owe its existence to the inability of the observer to completely specify the state of the system with infinite precision?}\\
\\
To put it differently: presumably, a monkey is more ignorant, as compared to a human physicist, about the complete set of degrees of freedom characterizing e.g. the falling of a glass of wine. Yet, the human is more certain about what will happen to the glass of wine as it falls. The directionality of time does not seem less clear to the human despite having more precise information about the state of the glass of wine. This contradicts the logical consequence of the second law as stated above within the subjective interpretation of probability. From the subjective point of view, the human is less limited than the monkey and therefore should lose information at a lesser rate. Hence, the monkey would have a greater sense of time's arrow. Thus, the subjectivist must conclude that either the monkey does have a much greater sense of the flow of time, or the directionality has nothing to do with the concept of entropy and the second law of thermodynamics. It is thus a matter of personal taste on which ideas should be kept as foundational and which ideas should be discarded. If the subjective school of thought is abandoned, then there is no incompatibility between the idea of a fundamental character for time's arrow and the second law of thermodynamics.

\subsection{Conclusion}

In the contemporary formulation of the thermodynamic arrow of time, as understood within the theory of statistical mechanics, the concept of entropy is employed to give an argument for the apparent directionality in time. It is a common belief that time's arrow is a fact following directly from the tendency of systems to evolve towards a statistical equilibrium where its entropy is at its maximum value. At the same time, a common interpretation of the concepts of probability and entropy in statistical mechanics is that they appear due to the lack of precise knowledge by the observer about the system under study. In this picture, probability and entropy are subjective concepts that depend both on the characteristics of the system and the observer.

In this article, it has been argued that the subjective information-theoretical interpretation of probability and entropy is incompatible with the philosophy that time's arrow is a fundamental property of Nature. For the subjectivist to uphold the idea that time's arrow is fundamental, it is necessary to abandon the idea that its origin has anything to do with the concept of entropy and the second law of thermodynamics.

\end{document}